# Chiral dual spin currents field-free perpendicular switching by altermagnet RuO$_2$

Gengchen Meng[1,2], Qi Sun[3], Zhicheng Xie[4], Yumin Yang[1,2], Yu Zhang[1,5], Na Lei[3], Dahai Wei[1,2*]

[1]State Key Laboratory of Semiconductor Physics and Chip Technologies, Institute of Semiconductors, Chinese Academy of Sciences, Beijing 100083, China
[2]Center of Materials Science and Optoelectronics Engineering, University of Chinese Academy of Sciences, Beijing 100190, China
[3]Fert Beijing Institute, MIIT Key Laboratory of Spintronics, School of Integrated Circuit Science and Engineering, Beihang University, Beijing 100191, China
[4]Southwest Institute of Technical Physics, Chengdu, 610041, China
[5]School of Industry-education Integration, University of Chinese Academy of Sciences, Beijing 100049, China

**Abstract**

Conventional spintronic mechanisms, such as spin-transfer and spin-orbit torques based on the spin current, rely on breaking time-reversal symmetry to manipulate magnetic moments. In contrast, for spatially separated dual spin currents, the time-reversal-invariant vector chirality emerges as a critical factor governing magnetization dynamics. Here, we investigate field-free perpendicular magnetization switching in an altermagnet RuO$_2$/ferromagnet/heavy metal Pt trilayer, driven by chiral dual spin currents (CDSC). We demonstrate that the chirality of these dual spin currents acts as the deterministic role in breaking out-of-plane symmetry. Leveraging the intrinsic spin-splitting effect of the d-wave altermagnet to generate an x-polarized spin component, the interplay of non-collinear spin currents from two adjacent layers induces a helical magnetic texture within the intermediate layer. The resulting intralayer exchange coupling manifests as an effective in-plane magnetic field, facilitating deterministic switching. This distinct physical picture, validated by switching measurements and micromagnetic simulations, reveals that the switching polarity is dictated by chirality rather than charge current polarity. Characterized by the novel symmetry and low power consumption, CDSC offers a promising paradigm for next-generation high-performance spintronic architectures.

# 1. Introduction

Manipulating magnetic order via spin currents lies at the very heart of spintronics. This fundamental concept has underpinned a host of physical phenomena and technological advances, including spin-transfer torque[1], spin-orbit torque[2], magnetic random access memories[3,4], and spin-torque nano-oscillators[5]. Acting as an external excitation, a spin current interacts directly with magnetic moments, driving rotation or precession away from equilibrium. While conventional strategies have focused on tailoring material properties[6,7], external fields[8], and geometries[9] to modulate these dynamics, while fundamentally engineering the intrinsic structure of the spin current itself also offers a pathway to unprecedented innovations. Typically, magnetization dynamics are governed by the spin-current polarization vector $\vec{\sigma}$ and the spin-current density $J_S$. Since $\vec{\sigma}$ is a time reversal breaking axial vector, inverting $\vec{\sigma}$ under time reversal reverses the magnetic dynamics. In contrast, the introduction of non-collinear dual spin currents creates a new degree of freedom characterized by a vector chirality $\vec{\sigma}_1 \times \vec{\sigma}_2$. Notably, this vector chirality remains invariant under time reversal. Consequently, chiral dual spin currents (CDSC) exhibit chirality dependent effects and symmetry properties distinct from their single spin current counterparts. Here, using CDSC-driven perpendicular magnetization switching as a paradigm, we elucidate the physical mechanism governing chirality-controlled switching polarity.

Realizing non-zero vector chirality CDSC necessitates a material architecture capable of generating non-collinear dual spin currents. While a $\sigma_y$-polarized spin current (transverse to the charge flow) is readily accessible via the spin Hall effect (SHE) in heavy metals[10,11], generating a $\sigma_x$ component is difficult through traditional methods. The recently emerged class of altermagnets offers a feasible solution[12,13]. Merging the characteristics of ferromagnets and antiferromagnets, altermagnetism combines spin-split band structures with antiparallel collinear magnetic order, giving rise to nontrivial phenomena such as the crystal Hall effect[14], nonlinear Hall effect[15], spin splitting effect (SSE)[16-18], and spin splitting torque (SST)[17,19]. Crucially, the SSE generates a spin current polarized parallel to the Néel vector $\vec{N}$ (Fig. 1(a)). Consequently, by tuning the angle between the charge current and specific crystal orientations, a distinct $\sigma_x$-polarized spin current component can be obtained. As illustrated in Fig. 1(b), we fabricated an altermagnet $RuO_2$(100)/ferromagnet [Ni\Pt]$_{3.5}$/heavy metal Pt (AM/FM/HM) trilayer heterostructure. We demonstrate chirality-controlled, field-free perpendicular magnetization switching. Both experimental observations and micromagnetic simulations align with the CDSC theoretical framework, confirming that the magnetization switching dynamics and chirality adhere to time-reversal symmetry.

## 2. Results and Discussion

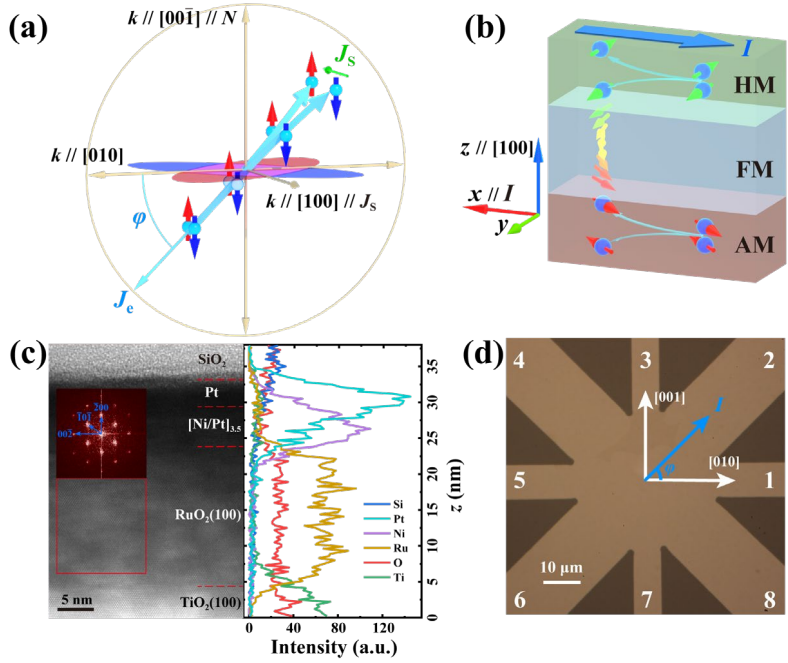

Figure 1. (a) SSE of $RuO_2(100)$. (b) Spin currents and magnetic moments in AM\FM\HM trilayer. (c) TEM image, selection FFT and EDS of trilayer sample. (d) 20 μm wide eight-terminal device

Consider the case of a $RuO_2(100)$ single crystal. The spin splitting effect (SSE) emerges specifically when the charge current aligns with the [100] or [010] crystal orientations. Given that the SSE-induced spin polarization is parallel to the $\vec{N}$ along the [001] orientation, injecting an in-plane current $I$ at an azimuthal angle $\varphi$ relative to the [010] orientation. As shown in Fig. 1(a), SSE generates a spin current characterized by both $\sigma_x$ and $\sigma_y$ in-plane polarization components:

$$J_{SRuO_2}\vec{\sigma}_{RuO_2} = \theta_{RuO_2}J_{RuO_2}\cos\varphi\,(\sin\varphi\,\hat{x} + \cos\varphi\,\hat{y}) \quad (1)$$

Concurrently, the SHE in the heavy metal Pt layer contributes a spin current defined by:

$$J_{SPt}\vec{\sigma}_{Pt} = -\theta_{Pt}J_{Pt}\hat{y} \quad (2)$$

denotes $\theta \equiv J_S/J_e$ and $J_e$ represents the charge current density. If simply consider the $\sigma_x$ component as the source of an effective in-plane field along the $x$-axis, while the $\sigma_y$ components from the two layers superpose (either reinforcing or partial cancelling), we derive the effective field and the critical switching current density by applying the macrospin model for field-assisted switching[20]:

$$H_{ex} = \chi J_{RuO_2}\cos\varphi\sin\varphi \quad (3)$$

$$J_C = \frac{e\mu_0 M_S t_{FM}}{\hbar \chi_{DL}}(H_K - \sqrt{2}|H_{ex}|) \propto \frac{H_K - \frac{\sqrt{2}}{2}\chi J_{RuO_2}|\sin 2\varphi|}{\theta_{Pt}J_{Pt} - \theta_{RuO_2}J_{RuO_2}\cos^2\varphi} \quad (4)$$

Here, $e$, $\mu_0$, $M_S$, $t_{FM}$, and $\hbar$ denote the elementary charge, vacuum permeability, saturation magnetization, ferromagnetic layer thickness, and reduced Planck constant, respectively. $\chi_{DL}$, $H_K$ and $\chi$ represent the efficiency of the damping-like effective field, effective anisotropic field, and efficiency of the $H_{ex}$, respectively. Experimental measurements indicate that the electrical conductivity of RuO$_2$ is approximately 0.8 times that of Pt layer, with $\theta_{RuO_2}$ being slightly smaller than $\theta_{Pt}$. Consequently, the derived $J_C$ exhibits minima at $\varphi \approx +45°$ and $-45°$. In contrast, for $\varphi = 0°$ and $\pm 90°$, $H_x$ vanishes, prohibiting deterministic switching.

We propose the following mechanism for the origin of $H_{ex}$ induced by the $\sigma_x$-polarized spin current. At the bottom interface of the ferromagnetic (FM) layer, the spin splitting torque (SST) of $\sigma_x$ tilts the magnetic moments from the perpendicular axis toward the *x*-axis. In systems with weak perpendicular magnetic anisotropy (PMA), this effect is pronounced, potentially forcing the bottom moments fully in-plane. Conversely, the top moments, dominated by the spin-orbit torque (SOT) from Pt, which exceeds the SSE strength, tend to align along the *y*-axis rather than *x*-axis. This competition results in a non-collinear magnetic configuration across the FM thickness [Fig. 1(b)], analogous to the exchange spring effect[21,22]; we term this phenomenon the spin torque spring effect. This non-collinearity introduces an *x*-component to the internal exchange coupling, manifesting as $H_{ex}$. The effect is likely amplified by the relatively weak ferromagnetic exchange coupling of Ni-Ni pairs in the [Ni/Pt] multilayers and the presence of Pt spacers. While phenomenologically similar to field-free switching in antiferromagnet/ferromagnet/heavy-metal (AFM/FM/HM) systems [23], the physical origins differ fundamentally: here, $H_{ex}$ arises dynamically from intralayer exchange coupling within the FM, whereas in AFM systems, it stems from static interfacial exchange bias.

Crucially, unlike a static external field, the current-induced $H_{ex}$ reverses sign when the current polarity is inverted, as does the $\sigma_y$-polarized spin current. Consequently, reversing the current polarity alone does not invert the switching direction. However, changing the sign of $\varphi$ reverses $H_{ex}$ while preserving the sign of the $\sigma_y$-polarization spin current, thereby inverting the switching direction. This behavior indicates that the switching polarity of this non-collinear dual spin current system remains invariant under time-reversal operation, distinct from conventional magnetization switching. To capture this unique symmetry, we introduce the vector chirality of dual spin currents $\vec{\kappa}$ as the parameter governing the switching polarity:

$$\vec{\kappa} \equiv \left(|J_{SRuO_2}|\vec{\sigma}_{RuO_2}\right) \times \left(|J_{SPt}|\vec{\sigma}_{Pt}\right) = -\frac{1}{2}\theta_{RuO_2}\theta_{Pt}|J_{RuO_2}J_{Pt}|\sin 2\varphi\, \hat{z} \propto \sin 2\varphi \quad (5)$$

The magnitude and sign of $\vec{\kappa}$ dictate the deterministic switching trajectory of the *z*-axis magnetization. Leveraging this specific symmetry, we demonstrate field-free perpendicular magnetization switching by modulating the sign of $\varphi$ and $\vec{\kappa}$, thereby validating the CDSC model. We note that this analysis assumes the perturbative approximation where the Pt spin current dominates, the regime where the two spin currents are comparable requires further meticulous theoretical treatment.

Detailed fabrication procedures and basic characterization data are provided in the Supplemental Material (S1–S4). Figure 1(c) presents cross-sectional transmission electron microscopy (TEM) image of the AM/FM/HM trilayer along the (010) orientation, with fast Fourier transform (FFT) analysis of RuO$_2$(100) layer and elemental mapping via energy dispersive X-ray spectroscopy (EDS). The structural analysis confirms that the RuO$_2$(100) layer grows epitaxially on the TiO$_2$(100) substrate. The heterostructures were patterned into eight-terminal Hall bar devices with channel widths of 10 or 20 µm using photolithography and argon ion etching, as depicted in Fig. 1(d). We probed the magnetic properties of the [Ni/Pt]$_{3.5}$ ferromagnetic layer using the anomalous Hall effect (AHE). As shown in Fig. 2(a), the system exhibits perpendicular magnetic anisotropy (PMA) at 300 K. Upon heating, the coercivity diminishes, marking a transition to an in-plane magnetic anisotropy (IMA) regime near 370 K, a trend attributed to the thermal degradation of the interfacial perpendicular exchange coupling at the Ni/Pt interface[24,25]. Subsequent measurements were performed at an ambient temperature of 300 K unless otherwise specified.

For current-induced continuous switching measurements, current was applied along the $\varphi = +45°$ (channel 6-2) and $\varphi = -45°$ (channel 4-8) directions, with voltage measured across the transverse channels (8-4 and 6-2), see Fig. 1(d). Under an applied current at $\varphi = \pm45°$, we observe a distinct horizontal shift in the magnetic hysteresis loops (Fig. 2(b)), indicating the emergence of a non-zero external magnetic field $\mu_0 H_z$ at the loop center. This symmetry breaking between the $\pm z$ magnetization states at zero external field favors a deterministic orientation, consistent with observations in other field-free switching systems[6,23,26,27]. We quantify this effect by extracting the out-of-plane bias field $\mu_0 H_b$ as a function of current density $J_{Pt}$ and $\varphi$, as shown in Fig. 2(c). While negligible at low currents, $\mu_0 H_b$ increases monotonically with $J_{Pt}$ in the range of 3~5 × 10$^{10}$ A/m$^2$ for $\varphi = \pm45°$. However, beyond a threshold of $J_{Pt} \approx 5 \times 10^{10}$ A/m$^2$, $\mu_0 H_b$ rapidly decays toward zero.

The non-monotonic behavior of $\mu_0 H_b$ arises from the interplay between spin torque dynamics and thermal effects. The initial rise stems from the increasing spin current density injected into the FM layer, which exert stronger torques, eventually overcoming magnetic anisotropy and damping to initiate precession. Concurrently, Joule heating becomes significant (see Supplemental Material S5). At $J_{Pt} = 4~5 \times 10^{10}$ A/m$^2$, the sample temperature reaches approximately 350 K, where the PMA is critically suppressed (Fig. 2(a)). This weakened anisotropy facilitates the canting of magnetic moments away from the easy axis under the spin torque, thereby amplifying $\mu_0 H_b$. Conversely, the subsequent collapse of $\mu_0 H_b$ is primarily driven by the transition toward a dominant IMA regime induced by intense Joule heating. Strong IMA confines the moments to the film plane, suppressing their response to out-of-plane perturbations from spin currents and magnetic fields. Furthermore, the reduction in saturation magnetization at elevated temperatures diminishes the effective spin torque. Simultaneously, the altermagnetic order and associated SSE in RuO$_2$ likely degrade as the temperature approaches the Néel temperature (approximate 400 K for bulk[28]).

The opposing signs of $\mu_0 H_b$ observed at $\varphi = +45°$ and $\varphi = -45°$ arise because the $\sigma_x$-polarized spin current inverts with the sign of $\varphi$ and $\vec{\kappa}$. As previously discussed, this inversion dictates both the switching polarity and the sign of $\mu_0 H_b$, as they share a common physical origin of CDSC: the synergy action of the SSE and the SHE. Consistent with our model, $\mu_0 H_b$ vanishes at $\varphi = 0°$ and $\pm90°$. Plotting the maximum bias field $\mu_0 H_{bm}$ against $\varphi$ reveals a dependency that closely follows the $\vec{\kappa}$ and $\sin 2\varphi$ relation, as shown in Fig. 2(d). This confirms that the vector chirality $\vec{\kappa}$ serves as a robust parameter for describing the $z$-axis symmetry breaking in this system.

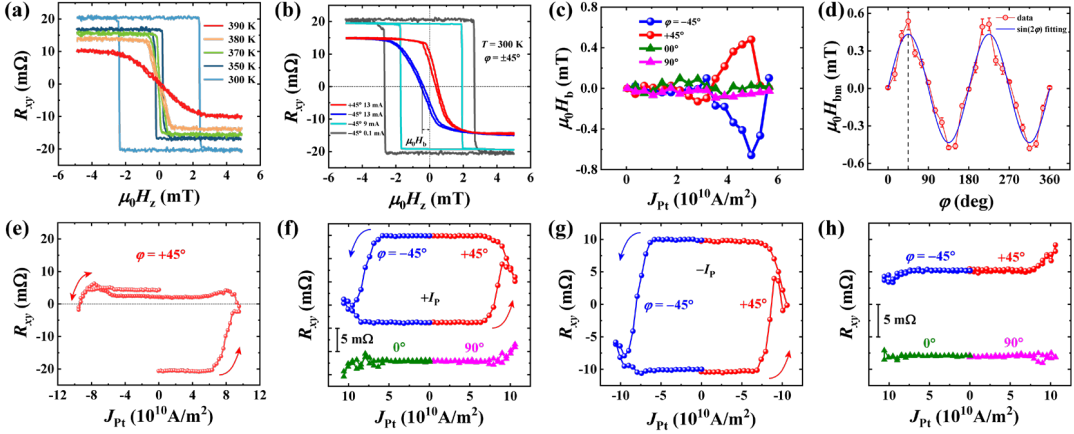

Figure 2. (a) AHE loops with different temperatures. (b) AHE loops with different current and $\varphi$. (c) $\mu_0 H_b$ versus $J_{Pt}$. (d) $\mu_0 H_{bm}$ versus $\varphi$. (e) Change current polarity scan. (f) Change $\varphi$ switching with fixed polarity $+I_P$ and (g) $-I_P$. (h) Switching of $RuO_2(110)$ trilayer sample.

To validate the proposed magnetization switching mechanism, we performed field-free switching measurements on the $RuO_2(100)$ sample. Figure 2(e) displays the result at a fixed angle of $\varphi = +45°$ while sweeping the current magnitude and polarity. Pulse width (PW) is 50 μs. Strikingly, regardless of the current polarity, the magnetic moments exhibit unidirectional switching from $-M_z$ to $+M_z$, failing to close the hysteresis loop. This behavior stands in sharp contrast to conventional spin-orbit torque switching, here, the switching polarity is effectively decoupled from the charge current polarity. This anomaly arises because inverting the charge current constitutes a time-reversal operation. Since the switching polarity is governed by the CDSC vector chirality $\vec{\kappa}$, a time-reversal invariant, the directional preference of the magnetization flip remains unchanged under current reversal.

We then fixed the current polarity $+I_P$ and modulated the sign of $\varphi$ to scan the switching curves, as shown in Fig. 2(f). At $\varphi = +45°$ and $\varphi = -45°$, the switching polarities are inverted, which switching from $-M_z$ to $+M_z$ and $+M_z$ to $-M_z$, respectively. By alternating the sign of $\varphi$, a complete hysteresis loop is recovered with a switching ratio of approximately 50%. The critical switching current density is found to be $J_{CPt} \approx 8 \times 10^{10}$ A/m². This relatively low threshold is attributed to the system's weak PMA (see Supplemental Material S4), which gives the magnetic moments highly susceptible to spin-current perturbations. As expected from the CDSC switching model, no deterministic switching is observed at $\varphi = 0°$ and $90°$. Measurements performed with fixed negative polarity $-I_P$ yield nearly identical switching behavior to the $+I_P$ case, as shown in Fig. 2(g), further corroborating the theoretical model.

To be carefully, we conducted control experiments on $RuO_2(110)$ sample, where $\varphi = 0°$ corresponds to current flow along the $[1\bar{1}0]$ orientation. In stark contrast to the (100) case, no deterministic switching was observed for any angle $\varphi = \pm45°$ or $\varphi = 0°$ and $90°$, as shown in Fig. 2(h). This result validates the crystallographic requisite of the mechanism: the SSE of $RuO_2$ dictates, spin polarization emerges only when charge currents collinear with the [100]/[010] axes. Consequently, in-plane currents in

RuO$_2$(110) fail to generate the necessary SSE and CDSC. Anomalous Hall data for the RuO$_2$(110) control sample are detailed in Supplemental Material S6.

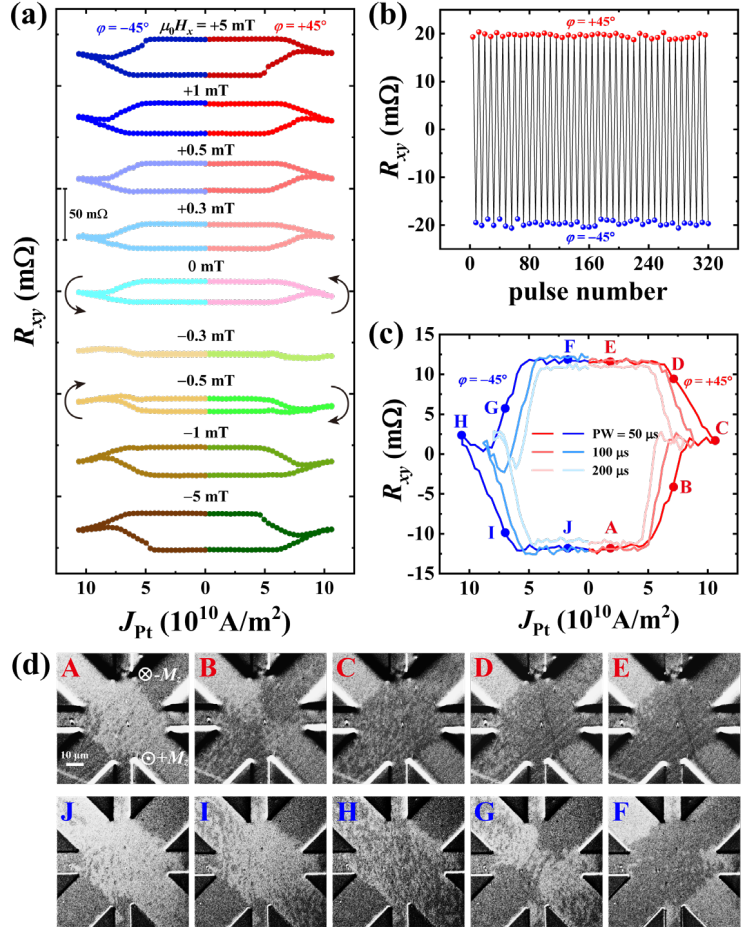

Figure 3. (a) Switching curves with $\varphi = +45°$ and $\varphi = -45°$ varying $\mu_0 H_x$. (b) consecutive switching cycles with $\varphi = +45°$ and $\varphi = -45°$. (c) Switching curves varying PW. (d) MOKE switching images.

Our theoretical model postulates that the SSE induces an in-plane effective field $H_{ex}$ along the $x$-axis, which acts in concert with the $\sigma_y$-polarized spin current to enable field-free perpendicular switching. To corroborate this mechanism and quantify $H_{ex}$, we measured the magnetization switching loops under varying external in-plane magnetic fields $\mu_0 H_x$, with $\varphi = +45°$ and $\varphi = -45°$, respectively, in Fig. 3(a). A distinct compensation point is observed at $\mu_0 H_x = -0.3$ mT, where switching is effectively suppressed. For fields diverging from this value $\mu_0 H_x > -0.3$ mT versus $\mu_0 H_x < -0.3$ mT, the switching polarity inverts. Furthermore, as the net effective field $|\mu_0 H_x + 0.3\,\text{mT}|$ increases, the switching ratio improves and the critical current density $J_{CPt}$ decreases. This behavior implies that the external field compensates for an intrinsic CDSC-induced field of $H_{ex} \approx 0.3$ mT.

To assess the robustness of the switching and validate the feasibility of the varying $\varphi$ measurement setting, we executed a durability test comprising 80 consecutive switching cycles. The sequence involved alternating pulse groups $\varphi = +45°$ and $\varphi = -45°$, respectively. Each group containing four pulses with $J_{Pt} = 8 \times 10^{10}$ A/m$^2$, a 50 μs PW, and a 60 s interval. The magnetic state was read out 30 s after each pulse group train via the AHE voltage across channels 7-3, with the probe current applied to channels 5-1 (Fig. 1(d)). As shown in Fig. 3(b), the magnetization switches reliably to $-M_z$ and $+M_z$ following the $\varphi = +45°$ and $\varphi = -45°$ pulse groups, respectively, demonstrating stable and deterministic field-free switching.

We further carry out the role of Joule heating effect. As illustrated in Fig. 3(c), $J_{CPt}$ decreases monotonically with increasing pulse width (PW), indicating that Joule heating reduces the PMA energy barrier, thereby facilitating switching. Switching process is directly visualized via magneto optical Kerr effect microscopy (MOKE), as shown in Fig. 3(d), where the magnetic domain evolution corresponds to the transport data (points A–F) in Fig. 3(c), which shows a clear light and dark contrast changing loop, confirming the nature of the magnetic switching. We observed that at $J_{Pt} = 8{\sim}10 \times 10^{10}$ A/m$^2$, the switching curve became relatively flat near $R_{xy} = 0$. This is because the larger $J_{Pt}$ causes the magnetic moment to completely turn within the plane and cannot be switched between $\pm z$ directions. It can be seen that the MOKE images at points C and H are both in the demagnetized state of multi-domains.

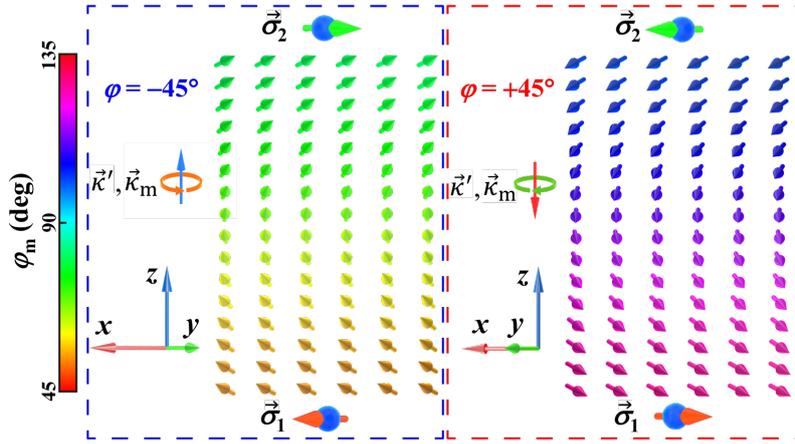

Figure 4. CDSC-induced helical magnetic structures.

To visualize the helical distribution of magnetic moments driven by the CDSC, we performed micromagnetic simulations using Mumax3.0 for intermediate magnetic structure of FM layer with $\varphi = +45°$ and $\varphi = -45°$, respectively, as shown in Fig. 4 (simulation details in Supplemental Material S7). The intermediate state refers to the equilibrium state when the CSDC is applied. The simulation reveals that the magnetic moments relax into a one-dimensional non-coplanar helical texture along the $z$-axis. The rotation of this helix is dictated by the sign of the angle $\varphi_{12}$ between the spin currents $\vec{\sigma}_1$ and $\vec{\sigma}_2$, which shares the same polarity as $\vec{\kappa}$ and $\vec{\sigma}_1 \times \vec{\sigma}_2$. Consequently, the azimuthal angle $\varphi_m$ of the in-plane component of local magnetization $m_{xy}$ acquires a gradient $\frac{\partial \varphi_m}{\partial z}$, which is determined by $\vec{\kappa}$. For a fixed $\varphi$, the intermediate local $z$-components $m_{zi}$ maintain a uniform sign across the layer, yielding a non-vanishing net magnetization $M_{zi}$ aligned with the polarity of $\vec{\kappa}$, corroborating our switching model. We further characterize this configuration via the magnetic vector chirality, defined as

$$\vec{\kappa}_{\mathrm{m}} \equiv \int_0^{t_{\mathrm{FM}}} m_{xy}^2 \frac{\partial \varphi_{\mathrm{m}}}{\partial z} dz \qquad (6)$$

Since the gradient $\frac{\partial \varphi_{\mathrm{m}}}{\partial z}$ is governed exclusively by the CDSC, the resulting magnetic chirality $\vec{\kappa}_{\mathrm{m}}$ aligns strictly with the chirality $\vec{\kappa}$ of the CDSC.

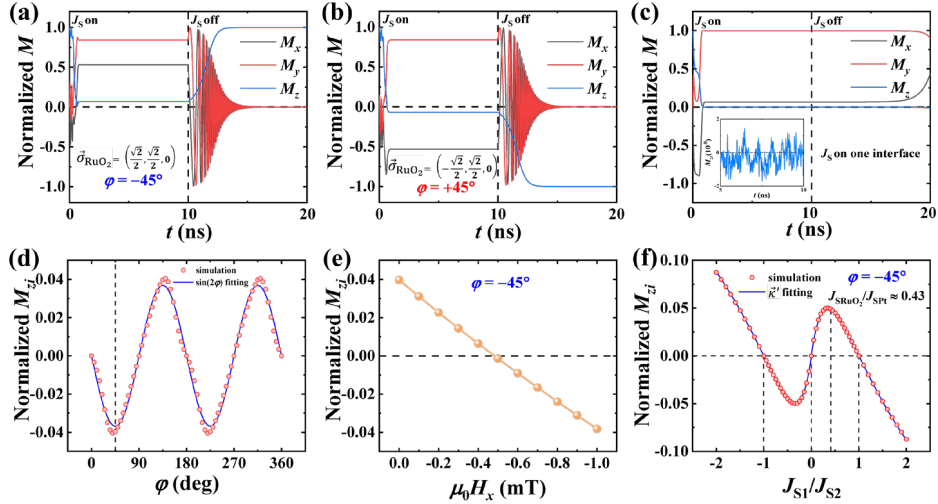

Figure 5. (a)(b) CDSC switching dynamics for $\varphi = +45°$ and $\varphi = -45°$, respectively. (c) magnetic moment dynamics induced by dual spin currents on the single interface. (d-f) The intermediate $M_{zi}$ versus (d) $\varphi$ (e) $\mu_0 H_x$ (f) $J_{S1}/J_{S2}$.

Micromagnetic simulations further corroborate the CDSC-driven switching mechanism. Upon removal of the CDSC, the magnetization relaxes to deterministic final states of $M_z = +1$ and $-1$ for $\varphi = +45°$ and $\varphi = -45°$, respectively, as shown in Fig. 5(a)(b). In contrast, applying dual spin currents to a single interface yields no deterministic switching (Fig. 5(c)), inset shows the intermediate $M_{zi}$. This result demonstrates the spatial separation of dual spin currents is requisite for switching. This texture-dependent phenomenon is inaccessible to macrospin models. We identify the $M_{zi}$ as a reliable predictor for the switching direction and strength, as it shares the sign of the final state $M_{zf}$. (See Supplemental Material S7 for conventional single spin current switching simulation and simulation codes).

Consistent with the bias field $\mu_0 H_{\mathrm{bm}}$ (Fig. 2(d)), $M_{zi}$ scales with the vector chirality $\vec{\kappa} \propto \sin 2\varphi$, as shown in Fig. 5(d), confirming $\vec{\kappa}$ as the governing factor for switching. Furthermore, $M_{zi}$ varies linearly with the external field, as shown in Fig. 5(e), vanishing at $\mu_0 H_x \approx -0.5$ mT, which is similar with the experimental behavior in Fig. 3(a). However, varying the spin currents ratio $J_{S1}/J_{S2}$ reveals that switching vanishes $M_{zi} = 0$ when $J_{S1}/J_{S2} = \pm 1$, which is different from the $\vec{\kappa}$-based prediction, as shown in Fig. 5(f). Here, the simple chirality $\vec{\kappa}$ argument requires modification. The internal exchange field $\vec{H}_e$ in FM orients between $\vec{\sigma}_1$ and $\vec{\sigma}_2$. Consequently, respect to the $\vec{\sigma}_1$ and $\vec{\sigma}_2$, the directions of $\vec{H}_e$ at the two surfaces oppose with each other. The opposite switching polarities $(\vec{H}_{e1} \times \vec{\sigma}_1)$ and $(\vec{H}_{e2} \times \vec{\sigma}_2)$ have

$$(\vec{H}_{e1} \times \vec{\sigma}_1) \cdot (\vec{H}_{e2} \times \vec{\sigma}_2) < 0 \tag{7}$$

leading to a cancellation effect of the two interfaces switching, which diminishes as $J_{S1}/J_{S2} = \pm 1$. In our devices of $\varphi = \pm 45°$, the ratio $J_{SRuO_2}/J_{SPt} \approx 0.43$ sits near the maximum of $M_{zi}$, enabling prominent switching. To generalize the polarity criterion for arbitrary spin currents ratios, accounting for competitive two interfaces torques $\left(\frac{(|J_{S2}|-|J_{S1}|)}{(|J_{S1}|+|J_{S2}|)}\right)$, and the switching suppression by the trivial in-plane damping-like effective fields ($\propto (|J_{S1}|\vec{\sigma}_1 + |J_{S2}|\vec{\sigma}_2)$), we propose a modified vector chirality

$$\vec{\kappa}' \equiv C \frac{(|J_{S2}| - |J_{S1}|)}{(|J_{S1}| + |J_{S2}|)} \frac{\vec{\kappa}}{||J_{S1}|\vec{\sigma}_1 + |J_{S2}|\vec{\sigma}_2|} = C \frac{(|J_{S2}| - |J_{S1}|)}{(|J_{S1}| + |J_{S2}|)} \frac{|J_{S1}J_{S2}|(\vec{\sigma}_1 \times \vec{\sigma}_2)}{\sqrt{J_{S1}^2 + J_{S2}^2 + 2J_{S1}J_{S2}\cos\varphi_{12}}} \tag{8}$$

where $C$ is a FM layer dependent constant. This vector chirality $\vec{\kappa}'$ provides a unified description of the switching polarity and accurately fits the simulated dependence of $M_{zi}$ on $J_{S1}/J_{S2}$, as shown in Fig. 5(f).

The distinct symmetry of the CDSC field-free switching allows us to rule out alternative mechanisms, such as interface exchange bias[23], in-plane composition gradients[29], structural asymmetry[9], tilted anisotropy[30], or $\sigma_z$ spin currents[26]. Unlike those scenarios, where switching typically reverses with current polarity, the process here is intrinsically controlled by the chirality $\vec{\kappa}'$ of the CDSC.

## 3. Conclusion

In summary, we demonstrate field-free perpendicular magnetization switching by exploiting the CDSC, which generated via the SSE in altermagnet $RuO_2$ and the SHE in heavy metal Pt. Our theoretical model reveals that the CDSC modulates the internal exchange coupling within the ferromagnetic layer, inducing an effective field $H_{ex} \approx 0.3$ mT. The observed perpendicular bias field $\mu_0 H_b$, which varies with current density and orientation, is quantitatively captured by this model when accounting for thermally induced magnetic variations. Robust switching occurs at $\varphi = \pm 45°$ pulse currents, with a small critical current density of $J_{CPt} \approx 8 \times 10^{10}$ A/m$^2$, while control experiments rigorously exclude alternative mechanisms. We establish the modified vector chirality $\vec{\kappa}'$ as the parameter governing switching polarity, which is able to explain well the switching process associated with $\varphi$ and $J_{S1}/J_{S2}$. These findings also confirm the SSE as a hallmark signature of altermagnetism. Furthermore, the CDSC-based architecture offers a novel paradigm for the design of next-generation spintronic devices.

**Acknowledgments**

This work was supported by the CAS project for Yong Scientists in Basic Research (Grant No. YSBR-030), the National Natural Science Foundation of China (Grant No.

12474102).

**Data Availability**

The data generated and analyzed here are available from the corresponding authors upon reasonable request.